# LOW EMITTANCE GROWTH IN A LEBT WITH UN-NEUTRALIZED SECTION*

L. Prost[#], J.-P. Carneiro, A. Shemyakin, Fermilab, Batavia, IL 60510, USA

*Abstract*

In a Low Energy Beam Transport line (LEBT), the emittance growth due to the beam's own space charge is typically suppressed by way of neutralization from either electrons or ions, which originate from ionization of the background gas. In cases where the beam is chopped, the neutralization pattern changes throughout the beginning of the pulse, causing the Twiss parameters to differ significantly from their steady state values, which, in turn, may result in beam losses downstream. For a modest beam perveance, there is an alternative solution, in which the beam is kept un-neutralized in the portion of the LEBT that contains the chopper. The emittance can be nearly preserved if the transition to the un-neutralized section occurs where the beam exhibits low transverse tails. This report discusses the experimental realization of such a scheme at Fermilab's PXIE, where low beam emittance dilution was demonstrated.

## SCHEME OF LEBT WITH UN-NEUTRALIZED SECTION

A Low Energy Beam Transport (LEBT) line in a modern ion accelerator typically connects an ion source (IS) to a Radio-Frequency Quadrupole (RFQ). Typical designs (e.g.: [1]) include 2 or more solenoidal lenses for focusing and rely on transport with nearly complete beam space charge neutralization over the entire length of the LEBT.

Reasoning and realization limitations for implementing a scheme where part of the LEBT is un-neutralized are discussed in some detail in Refs. [2, 3]. Major elements of the idealized scheme can be summarized as follows.

At the IS, the vacuum pressure is by default high, and the beam is nearly fully neutralized (Fig. 1).

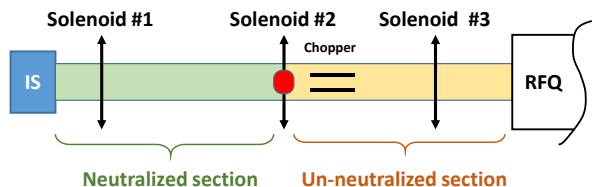

Figure 1: Transport scheme schematic.

The transition to un-neutralized transport is achieved by the combination of a potential barrier (red oval on Fig. 1) that confines neutralizing particles upstream and a clearing electric field (e.g.: DC offset on one of the chopper plates) that sweeps them out of the beam path downstream. In addition, a low vacuum pressure is maintained between the potential barrier and the RFQ to limit the rate at which neutralizing particles are created.

For a modest beam perveance, the main practical restriction for un-neutralized transport is emittance growth due to space charge non-linearities. To minimize this effect, one may consider a beam line designed with the following attributes:

- An ion source optimized to generate a uniform spatial density distribution;
- A completely neutralized beam transport from the ion source through Solenoid #1;
- At the image plane of the first focusing element, the distribution becomes again uniform. Neutralization is interrupted at this location;
- The phase advance over the remaining length of the LEBT is kept low. Hence the beam distribution stays close to uniform, and the emittance growth is suppressed.

## REALIZATION

The PXIE ion source delivers an H$^-$ beam of up to 10 mA DC at 30 keV. The LEBT nominal mode of operation is DC, However, for commissioning purposes, the LEBT is required to be able to provide a wide range of duty factors, which can be adjusted by varying the pulse length and/or pulse frequency. On the other hand, the elaboration of the LEBT transport scheme started with the idea of maintaining good vacuum in the RFQ, which the proposed scheme makes possible while avoiding long transient times due to space charge neutralization when pulsing.

A layout of the PXIE beam line before installation of the RFQ is shown on Figure 2. It consists of an H$^-$ Volume-Cusp Ion Source [4], 3 solenoids, a set of 4 transverse radiation-cooled scrapers (installed as temporary diagnostics between solenoids #1 & #2), a chopping system, Electrically Isolated Diaphragms (EID) (water-cooled, except for EID #4), an electrically-isolated, water-cooled, movable vertical electrode assembly with 3 apertures, and current diagnostics [5]. The chopper assembly consists of a 1000 l/s turbo pump, an electrostatic kicker and an EID. The kicker has two plates: one is grounded (and electrically-isolated) and also serves as the beam absorber; the second is biased to -5 kV to deflect the beam to the absorber, and brought towards ground to pass the beam. Details about most of the components can be found in [6]. Note that a modulator was added to the IS extraction electrode circuitry, thus providing pulsing capability independent of the chopper.

To realize the neutralization pattern shown in Figure 1, EIDs #1 and #2 are biased to +50 V to prevent background ions from moving from one section of the LEBT to another, while positive ions are cleared in the last ~1 m of the beam line before the RFQ by applying -300V DC voltage to the kicker plate.

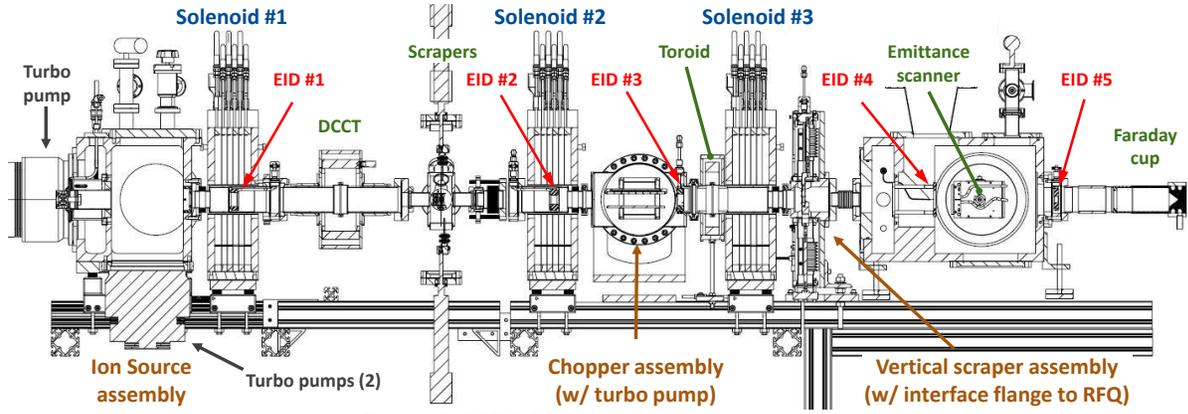

Figure 2: PXIE LEBT beam line (side view).

In the vicinity of the chopper, the vacuum pressure is of the order of $1\times10^{-7}$ Torr. In the IS extraction region, the vacuum pressure is at least 2 orders of magnitude higher.

In the simplest model, where neutralizing ions are confined longitudinally, the space-charge compensation builds up linearly until reaching an equilibrium determined by the balance between the radial loss of the compensating ions and their production. The time to reach such equilibrium (a.k.a. neutralization time) is determined only by the residual pressure and given by

$$\tau_{comp} = (n_{gas}\, \sigma_i\, v_p)^{-1}, \quad (1)$$

where $n_{gas}$ is the gas density, $\sigma_i = 1.5\times10^{-16}$ cm$^2$ [7] the ionization cross section of the $H_2$ gas, and $v_p$ the velocity of the $H^-$ ions. For the PXIE LEBT vacuum profile, the neutralization time given by Eq. (1) varies from microseconds near the ion source to milliseconds downstream of the chopper. Experimentally, the PXIE emittance scanner [8] provides time dependence data, which shows the effect of neutralization on the beam Twiss parameters over the pulse length. This effect was observed along a chopped pulse even with the clearing voltage on, indicating partial neutralization, likely due to the potential minimum at the beam waist between Solenoid #3 and the scanner. Note that when the beam is injected into the RFQ, such ion accumulation is eliminated because the beam waist is near the RFQ vanes, where the RF field cleans the ions out. To be closer to this case, most of the measurements described in this report were performed with a 50 μs pulse chopped out at the end of a 1.5 ms pulse formed by the IS modulator. The modulator pulse is long enough to allow reaching a steady state upstream of EID #2 while the chopped pulse is much shorter than the typical neutralization time downstream.

It should be noted that an aperture restriction at the exit of the ion source vacuum chamber significantly collimates the beam. For the nominal IS settings used to obtain 5 mA at the DCCT, we estimate that ~20% of the beam is scraped off. Simulations of the beam transmission through the 1st solenoid made with TRACK [9] agree to within 5%. In addition, they indicate that the beam emittance decreases noticeably, perhaps as much as 35% for this particular case.

## PROOF OF PRINCIPLE

Based on measurements and estimations that are outside the scope of this report, we believe that the neutralization profile in the experiments was reasonably close to the idealized step-function implied in Figure 1: ≥ 70% upstream of EID #2 and < 2% downstream. The following describes measurements and results that indicate that the scheme may have worked as intended.

### Beam distribution at the ion source

The Ion Source is commercially available and not necessarily optimized to deliver a beam with a uniform current density distribution. On the other hand, it seems natural to expect the beam current density distribution coming out of the ion source emitter to have sharp edges and be closer to being uniform rather than Gaussian. At the same time, the beam formation out of a plasma in a near thermal equilibrium must result in a Gaussian velocity distribution.

Information about the current density distribution can be extracted from phase portraits recorded with the emittance scanner installed near the ion source. Assuming the beam drifts in free space with no space charge, the phase space distribution can be propagated back toward the ion source with a simple linear coordinate transformation for each cell of the recorded distribution. Figure 3 shows an example of such back-propagation to the location of the ion source ground electrode, with cells distributed over 40 bins along the position coordinate.

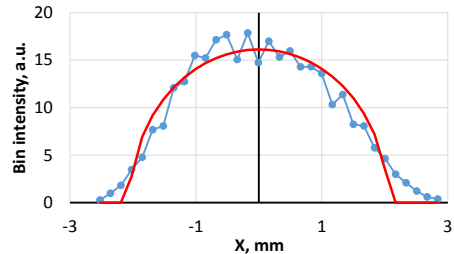

Figure 3: Comparison of a 1D current density distribution from a back propagated phase portrait (blue) and a uniform distribution fit (red).

The distribution displays features consistent with both a radially uniform distribution (1D projection, also plotted on Fig. 3) in the core and Gaussian tails. The collimation that takes place at the exit of the ion source vacuum chamber mostly removes the tails, producing at the image plane of Solenoid #1 a beam with a nearly uniform spatial distribution favourable to the transport scheme described here.

*Profile measurements*

Beam profiles (1D) were recorded upstream of EID #2 with scrapers (see Fig. 2). Figure 4a compares two distinctive cases with different Solenoid #1 currents, $I_{Sol\,\#1}$, but the same IS settings. The two curves differ significantly, showing similitudes with profiles corresponding to either a uniform or Gaussian current density distribution as judged by the sum of the squares of the differences between the data and profiles derived from ideal distributions with the same integral, 1st and 2nd moments. The best fit for each data set is shown on Fig. 4a (dashed lines).

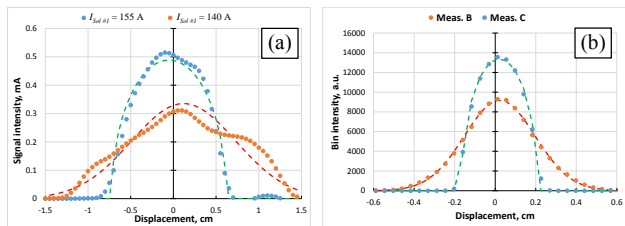

Figure 4: 1D beam profiles: (left, data) at the "scrapers"; (right, simulations) 20 cm downstream of EID #2. Dashed curves are fits, assuming a uniform (green) or a Gaussian distribution (red).

These profiles are reproduced well with TRACK simulations, which predict a similar behaviour downstream of EID #2 (Fig. 4b), where neutralization is assumed to be interrupted in the proposed transport scheme. Thus, for the value of $I_{Sol\,\#1}$ corresponding to Meas. C on Figure 4b, we may argue that the transition to un-neutralized transport in the experiment occurs with the beam having a current density distribution close to uniform.

*Emittance measurements*

Low emittance beams were measured at the end of the PXIE beam line under various biasing and focusing configurations (e.g.: [10]). Table 1 shows the results of 3 particular phase space measurements of interest: 1 at the exit of the ion source (A) and 2 downstream of solenoid #3 (B & C). For all cases, the ion source was tuned identically and the EIDs biasing configuration was the same. The data downstream is for a 50 μs chopped pulse, which, as mentioned previously, is much shorter than the neutralization time in that section of the beam line, hence a fair representation of an un-neutralized beam. As shown in the table, while focusing settings are significantly different, the measured Twiss parameters are nearly identical at the end of the beam line. Nevertheless, the measured emittances in all 3 cases are different.

Table 1: Phase space measurements results

|   | Sol. #1 [A] | Sol. #2 [A] | Sol. #3 [A] | $\varepsilon_n$ (rms) [μm] | α | β [m] |
|---|---|---|---|---|---|---|
| A | - | - | - | 0.19 | -3.5 | 0.6 |
| B | 154 | 187 | 223.5 | 0.25 | -8.9 | 2.2 |
| C | 143 | 158 | 240 | 0.16 | -8.2 | 2.2 |

We explain the decrease of the emittance between A & C by the scraping that takes place at the exit of the IS vacuum chamber. At the same time, the fact that, for B, the emittance is significantly larger than measured at the ion source exit (A) clearly shows that scraping alone does not necessarily lead to a beam with low emittance downstream. Conversely, space charge dominated transport does not necessary cause unacceptable emittance growth (C). Our interpretation is that measurement B corresponds to the case where the beam current density is not uniform near EID #2, while it is for measurement C (as illustrated by Fig. 4b). Note that it is merely coincidental that the value of $I_{Sol\,\#1}$ that leads to a uniform current density distribution in EID #2 is nearly identical to the one leading to a Gaussian distribution in Fig. 4a, and *vice-versa*.

Therefore, we believe that we have some reasonable evidence that the transport scheme with an un-neutralized section was realized and exhibit the properties enumerated in the first section, to within the uncertainties associated with a real accelerator.

## ACKNOWLEDGMENT

Authors recognize important contributions from B. Hanna, R. D'Arcy, V. Scarpine, and C. Wiesner in various experimental activities and data analyses at PXIE, and are grateful to the Accelerator Division supporting teams for making any of the measurements possible.

## REFERENCES

[1] L. R. Prost, FERMILAB-TM-2622-AD (2016), arXiv:1602.05488 [physics.acc-ph]
[2] A. Shemyakin, L. Prost, FERMILAB-TM-2599-AD (2015), arXiv:1504.06302 [physics.acc-ph]
[3] L. Prost, A. Shemyakin, *these proceedings*, MOPOY049
[4] www.d-pace.com
[5] R. D'Arcy et al., IPAC'15, Richmond, VA, USA (2015) MOPTY082
[6] L. Prost et al., IPAC'14, Dresden, Germany (2014) MOPRI086
[7] Y. Fogel et al., J. Explt. Theoret. Physics (USSR), **38** (1960) 1053
[8] R. D'Arcy et al., Nucl. Instr. and Methods in Phys. Res. A **815** (2016) 7-17
[9] B. Mustapha et al., PAC'05, Knoxville. TN, USA (2005) TPAT028
[10] L. Prost et al., IPAC'15, Richmond, VA, USA (2015) THPF126